\documentclass[10pt,twocolumn,letterpaper]{article}

\usepackage{cvpr}      
\newcommand{\cmark}{\ding{51}}%
\newcommand{\xmark}{\ding{55}}%
\usepackage{forest}
\usepackage{graphicx}
\usepackage{amsmath}
\usepackage{amssymb}
\usepackage{booktabs}
\usepackage{multirow}
\usepackage{svg}
\usepackage[utf8]{inputenc}
\usepackage[normalem]{ulem}
\useunder{\uline}{\ul}{}

\usepackage{algorithm}
\usepackage{algpseudocode}
\usepackage{pifont}
\usepackage{listings}
\usepackage{etoolbox}
\usepackage{lipsum}
\usepackage{mathtools}

\usepackage[pagebackref,breaklinks,colorlinks]{hyperref}
\definecolor{commentgreen}{rgb}{0.0, 0.27, 0.13}

\usepackage[pagebackref,breaklinks,colorlinks]{hyperref}

\usepackage[capitalize]{cleveref}
\crefname{section}{Sec.}{Secs.}
\Crefname{section}{Section}{Sections}
\Crefname{table}{Table}{Tables}
\crefname{table}{Tab.}{Tabs.}

\usepackage[subtle]{savetrees}
\usepackage{enumitem}
\def\cvprPaperID{16270} 
\def\confName{CVPR}
\def\confYear{2023}

\begin{document}


\title{DiCoM - Diverse Concept Modeling towards Enhancing Generalizability in Chest X-Ray Studies}
\author{
Abhijeet Parida$^1$
\qquad
Daniel Capellán-Martín$^{1,2}$
\qquad
Sara Atito$^3$
\qquad
Muhammad Awais$^2$\\
\qquad
María J. Ledesma-Carbayo$^2$
\qquad
Marius G. Linguraru$^{1,4}$
\qquad
Syed Muhammad Anwar$^{1,4}$\\
\\
\qquad
$^1$ Children’s National Hospital, Washington, DC, USA\\
{\tt\small \{pabhijeet,sanwar,mlingura\}@childrensnational.org} \\
$^2$ Universidad Politécnica de Madrid \& CIBER-BBN, Madrid, Spain\\
{\tt\small \{daniel.capellan,mj.ledesma\}@upm.es} \\
$^3$ CVSSP, University of Surrey, Surrey, UK\\
{\tt\small \{sara.atito,muhammad.awais\}@surrey.ac.uk} \\
$^4$ George Washington University, Washington DC, USA \\
{\tt\small \{syed.anwar\}@gwu.edu} \\
}
\maketitle

\begin{abstract}
   Chest X-Ray (CXR) is a widely used clinical imaging modality and has a pivotal role in the diagnosis and prognosis of various lung and heart related conditions. 
   Conventional automated clinical diagnostic tool design strategies relying on radiology reads and supervised learning, entail the cumbersome requirement of high-quality annotated training data. 
   To address this challenge, self-supervised pre-training has proven to outperform supervised pre-training in numerous downstream vision tasks, representing a significant breakthrough in the field. However, medical imaging pre-training significantly differs from pre-training with natural images (e.g., ImageNet) due to unique attributes of clinical images. 
   In this context, we introduce \textbf{Di}verse \textbf{Co}ncept \textbf{M}odeling (DiCoM), a novel self-supervised training paradigm that leverages a student teacher framework for learning diverse concepts and hence effective representation of the CXR data. 
   Hence, expanding 
   beyond merely modeling a single primary label within an image, instead, effectively harnessing the information from all the concepts inherent in the CXR. The pre-trained model is subsequently fine-tuned to address diverse domain-specific tasks. 
   Our proposed paradigm consistently demonstrates robust performance across multiple downstream tasks on multiple datasets, highlighting the success and generalizability of the pre-training strategy. 
   To establish the efficacy of our methods we analyze both the power of learned representations and the speed of convergence (SoC) of our models. 
   For diverse data and tasks, DiCoM is able to achieve in most cases better results compared to other state-of-the-art pre-training strategies. This when combined with the higher SoC and generalization capabilities positions DiCoM to be established as a foundation model for CXRs, a widely used imaging modality.
\end{abstract}

\section{Introduction}
Chest X-Rays (CXRs) are the preferred imaging modality for various clinical diagnostic tasks. This is supported by the fact that acquiring CXR is relatively cheap and the X-ray facilities are widely available. For pediatric cases, with CXRs, the exposure to radiation is lesser as compared to a computed tomography scan. Hence, CXRs are routinely used to triage and diagnose a variety of clinical conditions in pediatric population \cite{das2021comparison, UGALDE20211039}.  
Since the pandemic, CXRs have become the primary radiological screening method for Coronavirus disease (COVID-19), particularly for symptomatic or COVID-19 positive patients presented to the emergency department, as the Coronavirus disease manifests in an acute respiratory illness affecting the lungs. This has resulted in a large number of CXR studies performed in both adult and pediatric cases in the last few years. 
In general, with an increasing burden on healthcare facilities during the pandemic and beyond, 
CXR has proved very useful to triage COVID-19 and pneumonia inresource-constrained environments \cite{abo2021chest}. The success of this image-based diagnostic modality would be further enhanced by an expert computer-aided diagnosis system that can effectively process CXRs for pathology detection, thus managing to mitigate the growing workload for radiologists.

There are many deep learning methods presented in literature  for CXR analyses, yet very few are clinically adapted. Challenges towards the development and clinical translation of deep learning methods include the scarcity of quality labels, limited data for new or rare diseases (particularly for pediatrics), and lack of generalization for handling imaging data drift from different clinical sites and devices. Foundation medical image analysis models, similar to those trained on large-scale domain corpora in natural language processing \cite{gu2021domain}, could help address the challenges for image-based computer-aided diagnosis systems. 
However, for preserving patient privacy, clinical data needs to be securely stored, limiting availability, and access to large-scale shared clinical data sets. Further, it is harder to obtain pristine ground truth labels on clinical data to help with training fully supervised models. Still the amount of un-annotated digitized clinical data in medical centers is rapidly increasing. Patient outcomes would greatly benefit from the development of methods 
that learn from these large amounts of unused data. 

Within current deep learning architectures, vision transformers (ViT) trained with large-scale data achieve state-of-the-art performance in multiple vision-related tasks like classification, detection, and semantic segmentation \cite{han2022survey,parvaiz2022vision}. This success has led to the rapid development of vision foundation models for a variety of computer vision tasks\cite{azad2023foundational, ma2023foundation}. For medical imaging, the attention module in ViT can help identify appropriate features for the downstream domain-specific tasks, such as pathology classification or image segmentation, compared to the standard convolutional neural network (CNN) based architectures. Among these methods, masked image modeling strategies utilize large unlabeled data to facilitate effective self supervised learning (SSL) \cite{atito2022gmml, Kaiming2021mae, xie2022simmim}, hence 
providing optimal model weight initialization and convergence stability for multiple downstream tasks like classification and segmentation. 

Herein, we propose the Diverse Concept Modeling (DiCoM) strategy for domain-specific self-supervised pre-training using CXR data, 
thus ascertaining the feasibility of our proposed methodology for an automated CXR-based diagnosis and image analysis in the clinical domain. Particularly, we show that the learned representation is generalizable and successfully transferred to challenging small-scale and out of distribution unseen pediatric data with a significant performance in COVID-19 and pneumonia detection. We see performance gains in challenging cases such as multi-class classification and in a variety of other tasks related to CXR reads demonstrating the generlizability of our pre-training strategy.

\noindent \textbf{Our Contributions} are summarized as follows-
\setlist{nolistsep}
\begin{itemize}[noitemsep]
    \item We introduce DiCoM, diverse concept modeling strategy for self-supervised pre-training of deep learning model with a ViT backbone on chest X-Rays. 
    \item We conduct extensive downstream-tuning experiments on CXRs on both seen and unseen data distributions from public and private sources (to assess the relevance of the proposed method in practical clinical scenarios) for multi-class classification, binary classification, and semantic segmentation. 
    We establish the importance of pre-training on domain data for improved performance. 
    \item We are the first to study the effort required for the convergence of a large scale foundation model for clinical data and show that our training strategy is the most efficient. We also analyze the manifolds to evaluate and establish efficient representation learning. 
    \item To the best of our knowledge, we are the first to show that self-supervised pre-training on adult CXR helps learn useful representations for COVID-19 and multi-class pneumonia detection in pediatric chest scans, traditionally an underrepresented class in most data sets adding credence to the generalization of our model for unseen out of distribution data.   
\end{itemize}

\begin{figure*}[ht]
    \centering
    \includegraphics[width=0.90\linewidth]{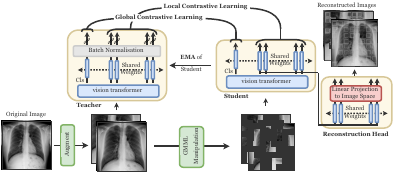}
    \abovecaptionskip=3pt
    \belowcaptionskip=0pt
    \caption{\textbf{Diverse Concept Modeling (DiCoM)} framework for self-supervised pre-training of vision transformers for chest X-ray image analysis.}
    \label{fig:DiCoMtransformer}
\end{figure*}

\section{Related Works} 

Earlier self-supervised pre-training methods utilized various pretext tasks, such as auto-encoding for reconstructing noisy or perturbed images~\cite{kramer1991nonlinear}, denoising auto-encoder approaches~\cite{vincent2008extracting}, predicting patch order~\cite{noroozi2016unsupervised}, color recovery from grayscale images~\cite{zhang2016colorful}, predicting missing regions via image inpainting~\cite{pathak2016context}, predicting rotation angles of images~\cite{gidaris2018unsupervised}, and adversarial loss for masked patches~\cite{trinh2019selfie}. Recent methods have shifted towards a contrastive learning framework~\cite{oord2018representation,wu2018unsupervised,hjelm2018learning,bachman2019learning,chen2020simple,NEURIPS2020_29539ed9,grill2020bootstrap}. The primary objective of contrastive learning is to minimize the distance between two augmented views of the same image while maximizing the distance between views from different images.

In recent years, driven by the popularity of vision transformers\cite{dosovitskiy2020image}, self-supervised learning approaches have adopted transformer backbones. The concept of masked image modeling for SSL pre-training has been explored in iGPT \cite{chen2020generative} and ViT \cite{dosovitskiy2020image}. These frameworks involve tasks like recovering quantized color space for patches using GPT and BERT tasks. While CNNs and ViTs in SSL have demonstrated potential, none have surpassed supervised pre-training, as exemplified by ViT trained on the JFT-300M Google internally labeled dataset \cite{dosovitskiy2020image}.

Self-supervised pre-training has been gaining appeal in the medical image domain. 
Inspired by the UNET architecture \cite{unet}, a transformer-based model was introduced for medical image analysis, particularly in segmentation \cite{unetr}. Employing a self-supervised pre-training strategy, the model demonstrated state-of-the-art performance on body organ segmentation challenge datasets. Another application involved a self-supervised contrastive learning strategy for COVID-19 prognosis \cite{sriram2021covid}, outperforming radiologists in predicting mortality for COVID-19 patients. Additionally, a self-supervised pre-training approach was applied to magnetic resonance imaging for knee abnormality classification \cite{atito2022sb}, highlighting the potential of such methods in limited data scenarios. This work represents a step forward in developing a generalizable foundational model for CXRs, showcasing the merits and feasibility of the proposed approach and its potential for improving patient outcomes.    

\section{Self-Supervised Pre-training for Medical Imaging}

Self supervised learning is an alternative to supervised learning, operating on the input data without the need for ground truth labels. The fundamental idea is to generate supervisory signals from the unstructured data by defining a pretext task, where a machine learning framework learns the underlying structure of the data to solve the task. To that end, SSL methods can be trained using a much larger data that are less affected by human labeling bias and noise within the data. 

Towards this, we introduce Diverse Concept Modeling- DiCoM, a general self-supervised vision transformer framework for CXRs. In Section \ref{sec:GMML}, we briefly summarise group masked model learning 
\cite{atito2021sit,atito2022gmml} and its adoption to medical domain. Following, in Section \ref{sec:Local_global}, we explain our proposed method of self-learning of data and class tokens conceptualization, with the incorporation of knowledge distillation \cite{hinton2015distilling}.
 
\subsection{Preliminary}
\label{sec:GMML}

Masked image modeling similar to the Masked Language Modeling (MLM) used in BERT~\cite{devlin2018bert}, was proposed in SiT \cite{atito2021sit} and employed in several recent vision works \cite{bao2021beit,atito2021mc,xie2022simmim}. Masked image modeling has different variations and group masked model learning \cite{atito2021sit} is one of them which proved effective when dealing with small datasets.The primary objective of group masked model learning is to corrupt a  ``significant'' segment of the visual input and restore them through model learning.  The idea is that by restoring the corrupted parts based on the overall context, the network implicitly grasps the concept of visual integrity. 

Various methods can be employed to corrupt a group of connected patches, such as replacing randomly selected patches with zeros, introducing noise, or substituting random patches with patches from other CXRs. In this study, we opted for simplicity and corrupted the input by replacing randomly selected groups of tokens with zeros. It's crucial to highlight that the groups of corrupted patches are selected randomly, i.e., patches with vital organs or background are manipulated with equal probability. This allows the network to generalize to unseen data, as demonstrated by the results for unseen datasets (Section \ref{sec:clf-unseen}).

To reconstruct the CXRs, we use ViT as a group masked auto-encoder. Analogous to auto-encoders, the network is trained to reconstruct the input image using the output tokens of the transformer. 
ViT processes a sequence of patches by tokenizing the input CXR $\mathbf{x} \in \mathbb{R}^{H \times W}$, where $H$ and $W$ represent the height and width of the CXR. The image is then flattened into $n$ non-overlapping 2D patches, of size $p \times p$ pixels.  

To pre-train the ViT, we manipulate the input CXR,  $\mathbf{x}$,  by randomly replacing 70\% of the image with zeros to obtain $\mathbf{\hat{x}}$. The reconstructed CXR, $\mathbf{\bar{x}}= D(E(\mathbf{\hat{x}}))$, is generated by inputting $\mathbf{\hat{x}}$ into the transformer encoder $E(.)$ and processing the output through a lightweight decoder $D(.)$. 
The decoder consists of three fully connected layers: the first two layers have 2048 neurons each, and the final layer has 256 neurons. Each layer is separated by a GeLU non-linearity. Transposed convolution operations are used to recover the CXR. Post pre-training, $E(.)$ is retained for downstream tasks and $D(.)$ is discarded. This decision is motivated by the fact that the $D(.)$ only learns CXR reconstruction, may not contribute significantly to the downstream tasks. In the reconstruction task, we use $\ell1$-loss between the original and the reconstructed image as shown in Equation \ref{eq:l1-pixel}. The loss is specifically computed for the corrupted pixel, similar to \cite{devlin2018bert,atito2022gmml}.

\begin{align}
\label{eq:l1-pixel}
\mathcal{L}_{\rm recons} &= \sum_i^H \sum_j^W \mathbf{M}_{i,j}\times | \mathbf{x}_{i,j} - \mathbf{\bar{x}}_{i,j} | 
\end{align}
where, $H\times W$ is the resolution of the CXR and $\mathbf{M}$ is a binary mask, with 1 corresponding to the manipulated pixels and 0 otherwise.

 \subsection{Local and Global Pseudo-Label Learning}
\label{sec:Local_global}
The DiCoM framework employs group masked model learning to grasp visual integrity. To enhance the concept further, we propose leveraging pseudo labels. Here, "pseudo" denotes labels from a teacher model during knowledge distillation. We suggest using self-knowledge distillation to learn local pseudo labels for individual tokens in the CXR and a global pseudo label for the entire CXR. Our hypothesis is that by using  multiple representations for an image might result in better representation learning, given the limited diversity in augmented views of CXRs. Hence we explore the learning of representation for each of the data tokens through knowledge distillation. 

In knowledge distillation, a student network is trained to mimic the output of a given teacher network, where $\theta$ and $\phi$ are the parameters of the student and the teacher networks, respectively. We employ an identical network architecture for both the student and the teacher, with the teacher acting as a momentum encoder~\cite{he2020momentum} for the student.An exponential moving average of the student weights is used to update the teacher network, following the rule: $\phi \leftarrow \lambda \phi + (1 - \lambda) \theta$.


In our case, the student network comprises of two components: A ViT backbone, followed by a projection head. The projection head consists of three fully connected layers: the first two with 2048 neurons and GeLU non-linearity each, and the final layer with 256 neurons. The output of the final layer is $\ell_2$ normalized and directly connected to a weight-normalized fully connected layer with 8192 neurons. It's important to note that the entire projection head is shared for both the class token and the data tokens. This shared configuration facilitates the transfer of semantics acquired in the class token to the patch tokens.  

To generate the pseudo label for the data tokens of the CXR image $\mathbf{x}$, the image is fed to the teacher network and the output tokens  are passed to the projection head to obtain $\mathbf{z}_t \in \mathbb{R}^{n \times K}$. The masked image $\mathbf{\hat{x}}$ is passed to the student network to obtain $\mathbf{z}_s \in \mathbb{R}^{n \times K}$, where $n$ is the number of tokens of the ViT and $K$, is the feature size of the final layer ($K=3092$). The objective is to align the output of the student network with that of the teacher network, utilizing the Kullback-Leibler(KL) divergence.


To prevent model collapse, an normalization step is introduced to avoid the model outputs becoming a trivial constant. The teacher's output goes through a batch normalization layer without adaptive parameters, using the moving average and standard deviation of the data, resulting in $\mathbf{\bar{z}}_t$. The output probability distributions are obtained as follows:
\vspace{-0.0cm}
   \begin{equation}
    p_t^{(i, j)} = \frac{\exp({\mathbf{\bar{z}}_t^{(i, j)}}/{\tau_t})}{\sum_{k=1}^K \exp({\mathbf{\bar{z}}_t^{(i, k)}}/{\tau_t})},
    \end{equation}
\begin{equation}
    p_s^{(i, j)} = \frac{\exp({\mathbf{{z}}_s^{(i, j)}}/{\tau_s})}{\sum_{k=1}^K \exp({\mathbf{{z}}_s^{(i, k)}}/{\tau_s})}
\end{equation}

\noindent   
where $\mathbf{\bar{z}}_t$ and $\mathbf{z}_s$ are the class logits for the teacher and the student. $p_t(i, .)$ and $p_s(i, .)$ are the probabilities corresponding to the $i^{\rm th}$ token output by the teacher and the student. Additionally, $\tau_t$ and $\tau_s$ are the temperature parameters for the teacher and the student. The softmax function includes a temperature parameter that controls the sharpness of the output probabilities. Sharpening is obtained by using a lower value for the temperature $\tau_{t}$ in the teacher compared to the student $\tau_{s}$. 
 
Our training objective is to match the output probability of the student $p_s$ with that of teacher $p_t$. We use the cross entropy measure for this task.
\begin{equation}
\mathcal{L}_{\rm l} = \sum_k^N \left( \sum_{i=1}^n T^k_i \times \sum_{j=1}^K  - p^k_t{(i, j)} \log p^k_s{(i, j)} \right),
\end{equation}
where $\mathbf{T}$ is a binary mask with 1 indicates a masked token and 0 otherwise. 


The network is trained to assign a class for each token in the image and distinguish it from other tokens. The idea of local contrastive loss ($\mathcal{L}_{\rm l}$) is similar to image reconstruction not in the pixel space, rather in the class space.Regarding global contrastive learning, we follow the same procedure as in the local contrastive learning, but focus on the class token to learn a global representation for the entire CXR. 
This task involves two augmented views. The goal is to match the student's class token from the first augmented view, $p_{s,1}(0, .)$, with the teacher's class token from the second augmented view, $p_{t,2}(0, .)$, and vice versa. The overall global contrastive learning loss($\mathcal{L}_{\rm g}$) is: 
\begin{equation}
\begin{split}
\mathcal{L}_{\rm g} &= \sum_k^N  \sum_{j=1}^K  - p^k_{t,1}{(0, j)} \log p^k_{s,2}{(0, j)}) \\
~~~~&+ \sum_{j=1}^C  - p^k_{t,2}{(0, j)} \log p^k_{s,1}{(0, j)}) 
\end{split}
\end{equation}

\subsection{Putting Together the DiCoM Framework} 

For a CXR, two augmented views are generated and fed into the teacher network. The masked version of the two views are passed to the student network. The output data tokens from the student network is used in both the reconstruction head and the contrastive learning projection head. The reconstruction loss, $\mathcal{L}_{\rm recons}$, is computed between the reconstructed images and the original images. Additionally, the local contrastive learning loss, $\mathcal{L}{\rm l}$, is determined by aligning the output of the data tokens from the student network with those from the teacher network. Subsequently, the global contrastive learning loss, $\mathcal{L}{\rm g}$, is calculated by matching the representation of the two augmented views corresponding to the class token from both the student and teacher networks. The overall loss, $\mathcal{L}$: 
\begin{equation}
    \mathcal{L} = \alpha_1  \mathcal{L}_{\rm recons} + \alpha_2 \mathcal{L}_{\rm l} + \alpha_3  \mathcal{L}_{\rm g}, 
\end{equation}
where $\alpha_1, \alpha_2$ \& $\alpha_3$ are the scaling factors of our diverse concept objective function, which are set to equal weights for simplicity in this study. We believe further improvements can be gained by optimizing the scaling factors or by incorporating  uncertainty  weighting \cite{kendall2018multi}. 

\begin{table*}[htbp]
\centering
\resizebox{0.95\textwidth}{!}{%
\begin{tabular}{@{}lcccccc@{}}
\toprule
\textbf{Dataset} & \textbf{Sample Size} & \multicolumn{1}{c}{\textbf{Classes / clinical conditions}} & \multicolumn{1}{c}{\textbf{Downstream task}} & \textbf{Pre-training} & \textbf{Fine-tuning} & \textbf{Testing} \\ \midrule
\textbf{COVIDxCXR-3}\footnote{\url{https://www.kaggle.com/datasets/andyczhao/covidx-cxr2}} \cite{Covidnet, pavlova2022covidx} & 30,384 & COVID-19, Pneumonia, Healthy & Binary Classification & \cmark & \cmark & \cmark \\
\textbf{CheXpert}\cite{Irvin2019} & 191,229 & 14 radiological observations & Binary Classification & \cmark & \cmark & \cmark \\
\textbf{ChestXray-14}\cite{Wang2017} & 112,120 & 14 radiological observations & - & \cmark & \xmark & \xmark \\
\textbf{SPR X-Ray}\cite{Kitamura2023} & 22,449 & Sex (male, female) & - & \cmark & \xmark & \xmark \\
\textbf{Shenzhen}\cite{Jaeger2014} & 662 & Tuberculosis, healthy & Segmentation & \cmark & \cmark & \cmark \\
\textbf{Montgomery}\cite{Jaeger2014} & 138 & Tuberculosis, healthy & Segmentation & \cmark & \cmark & \cmark \\
\textbf{Belarus}\cite{belarus} & 304 & Tuberculosis, healthy & Segmentation & \cmark & \cmark & \cmark \\
\textbf{MIMIC-CXR}\cite{Johnson2019physionet, Johnson2019arxiv} & 240,570 & 14 radiological observations & Binary Classification & \xmark & \cmark & \cmark \\
\multirow{2}{*}{\textbf{COVID-PEDs} (in-house)} & \multirow{2}{*}{199} & \multirow{2}{*}{COVID-19, healthy} & Binary Classification & \xmark & \cmark & \cmark \\
 &  &  & Convergence Analysis & \xmark & \xmark & \cmark \\
\textbf{Guangzhou Medical} \cite{Kermany2018} & 5,856 & Viral pneumonia, bacterial pneumonia, healthy & Multi-class Classification & \xmark & \cmark & \cmark \\
\multirow{2}{*}{\textbf{UNIFESP}\cite{unifesp}} & \multirow{2}{*}{2,480} & Chest, wrists, skull, thigh, lower leg, & \multirow{2}{*}{Representation Learning} & \multirow{2}{*}{\xmark} & \multirow{2}{*}{\xmark} & \multirow{2}{*}{\cmark} \\
 &  & Knee, elbow, feet, and finger  &  &  &  &  \\
 \bottomrule
\end{tabular}%
}
\abovecaptionskip=5pt
\caption{\textbf{Summary of Datasets} used for various tasks and experiments in this study.\vspace{-0.4cm} 
}
\label{tab:datasets}
\end{table*}

\section{Experimental Results}
\subsection{Datasets}

We aim to show that the domain-specific representation learned during the self-supervised pre-training stage using our proposed strategy (DiCoM) is generalizable and has the potential to solve frequent clinical tasks (classification and segmentation). To this end, we used a large-scale data comprising of 357,286 frontal CXR images for self-supervised pre-training, 
including pathologies such as COVID-19, pneumonia, tuberculosis, and clinical signs such as atelectasis, lung opacity, and cardiomegaly, among others. A brief summary 
is provided in Table \ref{tab:datasets}. Note that the initial CheXpert dataset contains a total of 224,316 CXR images (both frontal and lateral), of which our study uses 191,229 frontal CXR images.

Details including preprocessing steps and splits used for the pre-training, fine-tuning, and testing stages of the experiments are presented in Appendix \ref{sec:dataset}.
\subsection{Evaluation Strategies}

\noindent \textbf{Pre-Training:} We conducted training of the smaller ViT (ViT-S) model using several competitive SSL-based pre-training strategies, adhering to their suggested hyperparameters, for 250 epochs on the chest X-ray pre-training data (as described in Section 4.1 and Appendix A). The competing strategies employed for this study include the momentum contrastive method (MOCO-v3 \cite{he2020momentum}), knowledge distillation method (DINO \cite{caron2021emerging}), group masked methods (MAE \cite{Kaiming2021mae} and GMML \cite{atito2022gmml}), and SiT \cite{atito2021sit}.

 \noindent \textbf{Downstream Tuning:} After the self-supervised pre-training, task-specific layers for the downstream task are introduced. We hypothesize that the pre-training from unlabeled CXRs using our proposed strategy (DiCoM) learns excellent representation for task-specific downstream-tuning of network weights. We have further established this fact by objectively quantifying the learned representation (Section 4.6).  
 The fine-tuning for the downstream tasks is performed in a supervised manner for classification using DEiT \cite{pmlr-v139-touvron21a} and their suggested hyper-parameters for 100 epochs for a fair comparison. 
 To study the performance on segmentation and the impact of modification of the model architecture we use the UNETR framework \cite{unetr}. The UNETR using different weights from different pre-trained model are optimized for 200 epochs using their suggested hyper-parameters for 2D CXR images.  
 
\noindent \textbf{DiCoM Evaluation:} To assess various SSL strategies, we adhere to the guidelines outlined in \cite{balestriero2023cookbook}. In cases where labels for downstream tasks are available, we conduct both linear probing and fine-tuning. Linear probing involves appending a linear layer at the end of the frozen backbone to estimate the quality of the representation of the backbone \cite{balestriero2023cookbook}. Fine-tuning, on the other hand, involves re-training the entire network, including both the linear layer and the backbone, to gauge the total discriminative power of the entire model \cite{Kaiming2021mae}.

Extending our evaluation beyond classification tasks, we further assess SSL methods using tasks such as lung segmentation. Importantly, we delve into the evaluation of the speed of convergence (SoC) of downstream tasks in Section \ref{sec:low_resource} and explore the power of the learned representations in Section \ref{sec:rep_learn}.

\noindent \textbf{Metrics:} To compare different methods, we employ metrics that are agnostic to any thresholds and insensitive to class imbalances. Therefore, for binary classification, we utilize the Area Under the Precision-Recall (AUPR) curve, for segmentation, we use the Dice coefficient (equivalent to F1 score), and for clustering  to quantify learned representations, we rely on the Silhouette coefficient.

\subsection{Binary Classification}
\label{sec:clf_covid}
We conducted various downstream classification tasks to evaluate the efficacy of DiCoM across multiple datasets. 
\vspace{-0.2cm}

\subsubsection{Seen Distribution during Downstream Tuning}
\label{clf:seen}

We conducted both linear probing and fine-tuning experiments for COVID versus non-COVID classification on COVIDxCXR-3 data (refer to Table \ref{tab:results-covid}). This approach ensures that our reported results reflect performance on a seen data distribution, considering that COVIDxCXR-3 served as one of the datasets employed during the SSL pre-training phase. Our approach exhibits superior performance compared to alternative methods. Notably, we observe a positive AUPR difference of $\sim$5.2\% and $\sim$0.3\% during linear probing and fine-tuning, respectively. These findings suggest that the knowledge acquired through the self-supervised pre-training phase using DiCoM is more effectively transferred when adapting to domain-specific scenarios during the fine-tuning process.

We conducted both linear probing and fine-tuning experiments on another seen data distribution Chexpert,  using the training and testing data from CheXpert, adhering to the original data splits. Results are presented in Table \ref{tab:results-multi-bin}. Our findings indicate that, while DiCoM slightly trails behind other methodologies such as SiT, DINO, or MOCO in certain scenarios, it generally demonstrates competitive performance on CheXpert. Notably, we observe a performance improvement of $\sim$1.3\% and $\sim$0.5\% in AUPR curve through linear probing and fine-tuning, respectively.


\begin{table*}[htbp]\small
\centering
\resizebox{0.9\textwidth}{!}{%
\begin{tabular}{@{}l|cccccc|cccccc@{}}
\toprule
\multicolumn{1}{c|}{\multirow{3}{*}{\textbf{\begin{tabular}[c]{@{}c@{}}SSL \\ Technique\end{tabular}}}} & \multicolumn{6}{c|}{\textbf{COVIDxCXR-3}} & \multicolumn{6}{c}{\textbf{COVID-PEDs}} \\ \cmidrule(l){2-13} 
\multicolumn{1}{c|}{} & \multicolumn{3}{c|}{\textbf{Linear Probing}} & \multicolumn{3}{c|}{\textbf{Fine-tuning}} & \multicolumn{3}{c|}{\textbf{Linear Probing}} & \multicolumn{3}{c}{\textbf{Fine-tuning}} \\ \cmidrule(l){2-13} 
\multicolumn{1}{c|}{} & \textbf{ACC} & \textbf{AUPR} & \multicolumn{1}{c|}{\textbf{AUC}} & \textbf{ACC} & \textbf{AUPR} & \textbf{AUC} & \textbf{ACC} & \textbf{AUPR} & \multicolumn{1}{c|}{\textbf{AUC}} & \textbf{ACC} & \textbf{AUPR} & \textbf{AUC} \\ \midrule
\textbf{No Pre-training} & .855 & .948 & \multicolumn{1}{c|}{.936} & .971 & .997 & .997 & .605 & .681 & \multicolumn{1}{c|}{.667} & .684 & .820 & .772 \\
\textbf{MOCO-v3} & .935 & .989 & \multicolumn{1}{c|}{.986} & .984 & {\ul .999} & {\ul .999} & .658 & .735 & \multicolumn{1}{c|}{.678} & .737 & .847 & .808 \\
\textbf{DINO} & .978 & {\ul .998} & \multicolumn{1}{c|}{{\ul .998}} & .986 & {\ul .999} & {\ul .999} & {\ul .711} & .811 & \multicolumn{1}{c|}{.728} & .711 & .832 & .790 \\
\textbf{MAE} & .966 & .997 & \multicolumn{1}{c|}{.996} & \textbf{.993} & \textbf{1.00} & \textbf{1.00} & .684 & .815 & \multicolumn{1}{c|}{.750} & \textbf{.816} & \textbf{.917} & \textbf{.868} \\
\textbf{GMML} & .926 & .986 & \multicolumn{1}{c|}{.983} & .986 & {\ul .999} & {\ul .999} & .658 & .775 & \multicolumn{1}{c|}{.742} & {\ul .789} & .873 & .792 \\
\textbf{SiT} & {\ul .988} & \textbf{1.00} & \multicolumn{1}{c|}{\textbf{1.00}} & {\ul .992} & \textbf{1.00} & \textbf{1.00} & .632 & {\ul .821} & \multicolumn{1}{c|}{{\ul .772}} & {\ul .789} & .865 & .808 \\
\textbf{DiCoM (ours)} & \textbf{.989} & \textbf{1.00} & \multicolumn{1}{c|}{\textbf{1.00}} & .990 & \textbf{1.00} & \textbf{1.00} & \textbf{.763} & \textbf{.848} & \multicolumn{1}{c|}{\textbf{.783}} & .711 & {\ul .879} & {\ul .828} \\ \bottomrule
\end{tabular}%
}
\abovecaptionskip=5pt

\caption{\textbf{Binary classification} of  non-COVID-19 vs. COVID-19 on test sets of COVIDxCXR-3 (adult, seen) and COVID-PEDs (private data, pediatric, unseen). The highest scores are in bold, while the second best ones are underlined. Abbreviations: ACC, accuracy; AUPR, Area Under the Precision-Recall Curve; AUC, Area Under the Receiver Operating Curve.\vspace{-0.25cm}}
\label{tab:results-covid}
\end{table*}

\begin{table*}[htbp]\small
\centering
\resizebox{0.85\textwidth}{!}{%
\begin{tabular}{@{}l|cccccc|cccccc@{}}
\toprule
\multicolumn{1}{c|}{\multirow{3}{*}{\textbf{\begin{tabular}[c]{@{}c@{}}SSL \\ Technique\end{tabular}}}} & \multicolumn{6}{c|}{\textbf{CheXpert}} & \multicolumn{6}{c}{\textbf{MIMIC-CXR}} \\ \cmidrule(l){2-13} 
\multicolumn{1}{c|}{} & \multicolumn{3}{c|}{\textbf{Linear Probing}} & \multicolumn{3}{c|}{\textbf{Fine-tuning}} & \multicolumn{3}{c|}{\textbf{Linear Probing}} & \multicolumn{3}{c}{\textbf{Fine-tuning}} \\ \cmidrule(l){2-13} 
\multicolumn{1}{c|}{} & \textbf{ACC} & \textbf{AUPR} & \multicolumn{1}{c|}{\textbf{AUC}} & \textbf{ACC} & \textbf{AUPR} & \textbf{AUC} & \textbf{ACC} & \textbf{AUPR} & \multicolumn{1}{c|}{\textbf{AUC}} & \textbf{ACC} & \textbf{AUPR} & \textbf{AUC} \\ \midrule
\textbf{No Pre-training} & .869 & .956 & \multicolumn{1}{c|}{.775} & .842 & .982 & .894 & .797 & .895 & \multicolumn{1}{c|}{.664} & .817 & .911 & .740 \\
\textbf{MOCO-v3} & {\ul .884} & .983 & \multicolumn{1}{c|}{.916} & \textbf{.907} & {\ul .993} & {\ul .962} & .819 & .922 & \multicolumn{1}{c|}{.750} & .821 & .929 & .777 \\
\textbf{DINO} & \textbf{.888} & {\ul .990} & \multicolumn{1}{c|}{{\ul .948}} & .896 & .992 & .959 & .822 & .914 & \multicolumn{1}{c|}{.747} & .828 & .930 & .777 \\
\textbf{MAE} & .849 & .990 & \multicolumn{1}{c|}{.935} & .884 & .988 & .934 & .829 & .926 & \multicolumn{1}{c|}{.768} & {\ul .839} & .938 & {\ul .791} \\
\textbf{GMML} & .855 & .983 & \multicolumn{1}{c|}{.895} & .855 & .993 & .953 & .825 & .918 & \multicolumn{1}{c|}{.749} & .821 & .933 & .779 \\
\textbf{SiT} & .869 & \textbf{.992} & \multicolumn{1}{c|}{\textbf{.959}} & {\ul .900} & \textbf{.995} & \textbf{.972} & {\ul .838} & \textbf{.953} & \multicolumn{1}{c|}{\textbf{.817}} & .820 & {\ul .939} & .774 \\
\textbf{DiCoM (ours)} & .865 & .969 & \multicolumn{1}{c|}{.852} & .861 & .987 & .930 & \textbf{.843} & {\ul .948} & \multicolumn{1}{c|}{{\ul .802}} & \textbf{.840} & \textbf{.947} & \textbf{.807} \\ \bottomrule
\end{tabular}%
}
\abovecaptionskip=5pt
\caption{\textbf{Binary classification} of multiple diseases on the test set of CheXpert (seen) and MIMIC-CXR (unseen) datasets. Original test splits were used for each of the datasets. Binary classification is disease vs. healthy. The highest scores are in bold, while the second best ones are underlined. Abbreviations: ACC, accuracy; AUPR, Area Under the Precision-Recall Curve; AUC, Area Under the Receiver Operating Curve.\vspace{-0.25cm}}
\label{tab:results-multi-bin}
\end{table*}
\subsubsection{Unseen Distribution during Downstream Tuning} \label{sec:clf-unseen}
We also conducted experiments involving the fine-tuning and linear probing of each pre-trained model on unseen data to assess the transferability of knowledge acquired during the pre-training phase, even when dealing with distribution-shifted data. To this end, the models were tuned and tested on MIMIC-CXR, maintaining the original splits. We kept MIMIC-CXR data for this particular experiment and hence was not included in the pre-training stage. Table \ref{tab:results-multi-bin} presents the results, demonstrating that DiCoM achieves superior performance on MIMIC-CXR. Specifically, it emerges as the top-performing approach in fine-tuning and ranks second in linear probing (with AUPR as ranking metric). When compared to a fully supervised ViT-S model, DiCoM exhibits a performance improvement of $\sim$5.3\% and $\sim$3.6\% in AUPR for linear probing and fine-tuning, respectively.

These results underscore DiCoM's robust adaptability to unseen data, showcasing its effectiveness in enhancing model performance in scenarios where conventional fully supervised approaches would not perform as well.

\vspace{-0.25cm}
\subsubsection{Shifted Distribution during Downstream Tuning}

In addition to evaluating the model's adaptability to unseen data, we investigated its performance in handling out-of-distribution (OOD) datasets, i.e., when there is a shift in the data distribution, specifically in the context of pediatric cases, where radiological signs and disease detection pose unique challenges in standard clinical practice.

Here, our experiments focused on COVID-PEDs, a relatively small dataset (approximately 1,000 times smaller than MIMIC-CXR or CheXpert) of pediatric COVID-19 cases collected in-house. The results, outlined in Table \ref{tab:results-covid}, highlight DiCoM's consistent high performance in handling task-specific OOD datasets. Notably, when compared to a fully supervised ViT-S model, DiCoM demonstrates a performance improvement of 16.7\% and 5.9\% in AUPR for linear probing and fine-tuning, respectively, underscoring its efficacy in addressing challenges posed by distribution-shifted pediatric data.

\subsection{Multi-class Classification}

To further assess the effectiveness of the representation learning using DiCoM, we conducted experiments on a multi-class classification task using the Guangzhou pediatric dataset \cite{Kermany2018}. This dataset not only serves as unseen data, but also poses an OOD scenario, given its distinct pediatric characteristics.

Results are presented in Table \ref{tab:results-pneumonia}. Here, once more, DiCoM showcased exceptional performance, securing the top score among all SSL techniques during fine-tuning and ranking second only to DINO in the case of linear probing. Comparative analysis against a fully supervised ViT-S model revealed notable performance improvements for DiCoM, with gains of 18.6\% and 18.3\% in AUPR for linear probing and fine-tuning, respectively.

This reaffirms DiCoM's efficacy in capturing and leveraging meaningful representations, consistently outperforming fully supervised and other self-supervised approaches even in complex multi-class classification scenarios. The superior performance on the Guangzhou pediatric dataset further underscores the robustness and versatility of DiCoM in diverse medical imaging tasks adding credence to the generalization capabilities of DiCoM.

\begin{table}[htbp]
\centering
\resizebox{0.95\columnwidth}{!}{%
\begin{tabular}{@{}l|c|c@{}}
\toprule
\textbf{SSL Technique} & \textbf{Linear Probing (AUPR)} & \textbf{Fine-tuning (AUPR)} \\ \midrule
\textbf{No Pre-training} & 0.673 & 0.708 \\
\textbf{MOCO-v3} & 0.827 & 0.827 \\
\textbf{DINO} & \textbf{0.875} & {\ul 0.888} \\
\textbf{MAE} & 0.846 & 0.804 \\
\textbf{GMML} & 0.740 & 0.833 \\
\textbf{SiT} & 0.776 & 0.853 \\
\textbf{DiCoM (ours)} & {\ul 0.859} & \textbf{0.891} \\ \bottomrule
\end{tabular}%
}
\abovecaptionskip=5pt
\caption{\textbf{Multiclass classification accuracy} for healthy vs. viral pneumonia vs. bacterial pneumonia on the test set of the Guangzhou pediatric pneumonia dataset.}
\label{tab:results-pneumonia}
\end{table}

\subsection{Segmentation: Architecture Shift}\label{sec:lungsegmentation}
In addition to the tasks with vanilla vision transformer architecture, we explored the performance of the pre-trained weights embedded into alternative architectures. Specifically, we evaluated the pre-training method in the context of lung segmentation using the UNETR \cite{unetr} architecture. 
\begin{table}[htbp]
\centering
\resizebox{0.95\columnwidth}{!}{%
\begin{tabular}{@{}l|cc|cc|cc@{}}
\toprule
\multicolumn{1}{c|}{\multirow{2}{*}{\textbf{SSL   Technique}}} & \multicolumn{2}{c|}{\textbf{Shenzhen}} & \multicolumn{2}{c|}{\textbf{Montgomery}} & \multicolumn{2}{c}{\textbf{Belarus}} \\ \cmidrule(l){2-7} 
\multicolumn{1}{c|}{} & \textbf{DICE} & \textbf{HD\_95} & \textbf{DICE} & \textbf{HD\_95} & \textbf{DICE} & \textbf{HD\_95} \\ \midrule
\textbf{No Pretraining} & 0.914 & 9.62 & 0.956 & 8.89 & 0.859 & 21.40 \\
\textbf{MOCO-v3} & 0.917 & {\ul 7.50} & {\ul 0.964} & {\ul 5.64} & 0.873 & 16.51 \\
\textbf{DINO} & 0.956 & 8.20 & 0.958 & 7.24 & 0.866 & 19.43 \\
\textbf{MAE} & \textbf{0.960} & \textbf{6.97} & \textbf{0.966} & \textbf{4.90} & {\ul 0.877} & {\ul 16.19} \\
\textbf{GMML} & 0.952 & 8.30 & 0.955 & 7.23 & 0.837 & 21.84 \\
\textbf{SiT} & 0.957 & 7.80 & 0.952 & 9.31 & 0.868 & 18.61 \\
\textbf{DiCoM (ours)} & {\ul 0.958} & 7.90 & {\ul 0.964} & 8.06 & \textbf{0.885} & \textbf{15.97} \\ \bottomrule
\end{tabular}%
}
\abovecaptionskip=5pt
\caption{\textbf{Lung segmentation} performance for different pre-trained weights on various dataset using UNETR\cite{unetr} for segmentation of lung. Abbreviations: HD\_95, Hausdorff 95\% distance.\vspace{-0.3cm}}
\label{tab:results-seg}
\end{table}

The segmentation performance is showcased on the test sets of three datasets: Shenzhen, Montgomery, and Belarus. The segmentation results for these datasets, characterized by varying sizes, are summarized in Table \ref{tab:results-seg}. Across all datasets, it is evident that the SSL pre-training proves advantageous for enhancing Dice scores. 

In the context of larger datasets (Shenzhen \& Montgomery), MAE\cite{Kaiming2021mae} based pre-training surpasses other methods in terms of Dice scores, with our proposed method closely following. However, for smaller dataset like the Belarus TB dataset, DiCoM outperforms other pre-training strategies. These findings highlight the efficacy of pre-trained weights for tasks involving changes in architecture as well. Overall, DiCoM demonstrates superior performance compared to alternative approaches. A comprehensive qualitative comparison of lung segmentation by different methods is given in Appendix B.
\subsection{Learning Representations}\label{sec:rep_learn}

In addition to assessing various downstream tasks, we also conducted an evaluation of the ViT-encoder's performance as a feature extractor. Our specific focus was on understanding how well the features learned for CXRs compared to X-Rays (XRs) from other body parts. 

We employed the UNIFESP dataset\cite{unifesp}, encompassing CXRs and XRs of non-intersecting body parts such as wrists, skull, thigh, lower Leg, knee, elbow, feet, and finger. The goal was to achieve equal representation of both classes. Our hypothesis was that CXRs would reside on a different manifold compared to the other XRs. To quantify this hypothesis, we utilized the Rand Index (a measure of clustering accuracy) and Silhouette Coefficient (a measure of cluster tightness), assuming the presence of two clusters — one for CXRs and another for other XRs. The rationale being that a good feature extractor would exhibit a high Rand Index and Silhouette Coefficient.

\begin{table}[htbp]
\centering
\resizebox{0.75\columnwidth}{!}{%
\begin{tabular}{@{}lcc@{}}
\toprule
\textbf{SSL Technique} & \multicolumn{1}{l}{\textbf{Rand Index}} & \multicolumn{1}{l}{\textbf{Silhoutte Coefficient}} \\ \midrule
\textbf{MOCO-v3} & 0.500 & 0.132 \\
\textbf{DINO} & 0.591 & 0.052 \\
\textbf{MAE} & \textbf{0.779} & 0.314 \\
\textbf{GMML} & 0.665 & \textbf{0.372} \\
\textbf{SiT} & 0.502 & 0.052 \\
\textbf{DiCoM (ours)} & {\ul 0.758} & {\ul 0.358} \\ \bottomrule
\end{tabular}%
}
\abovecaptionskip=5pt
\caption{\textbf{Clustering metrics} to evaluate feature learning of CXRs versus other X-Rays. Rand index is a measure of accuracy of clustering and silhoutte coefficient is a measure of tightness of the cluster.\vspace{-0.25cm}}
\label{tab:results-manifold-viz}
\end{table}
The results in Table \ref{tab:results-manifold-viz} reveal that for the Rand Index, MAE outperforms others with an accuracy of 77.9\%, while DiCoM closely follows with 75.8\%. Regarding the proximity of samples within each cluster, the highest performance is observed in pre-training using GMML, yielding a Silhouette Coefficient of 0.372, and DiCoM comes second with 0.358. When combined effect of both these metrics (silhoutte coefficient and Rand index) is considered, DiCoM has learned a more effective representation due to highly accurate closed clustering of CXRs compared to other methods, establishing itself as a superior feature extractor for this modality.

The manifold visualizations for each of the methods, complementing the metrics presented in Table \ref{tab:results-manifold-viz}, are provided in Appendix \ref{section:appendix-manifold}.

\subsection{Downstream Tuning in Low Resource Settings and Clinical Translation}
\label{sec:low_resource}

In the majority of medical interventions globally, resources, both in terms of data and computational power, are often limited. The use of pre-trained methods proves advantageous in such scenarios, enabling computationally intensive tasks to be offloaded to large compute clusters located remotely. Ideally, a clinical site dedicated to a specific task should find it sufficient to train for a few iterations on a small dataset to achieve optimal results. This approach enhances the feasibility and efficiency of implementing medical interventions in environments with limited resources. Moreover, the need for fewer computational resources leads to models having a reduced carbon footprint.

\begin{figure}[ht]
    \centering
    \includegraphics[width=0.85\columnwidth]{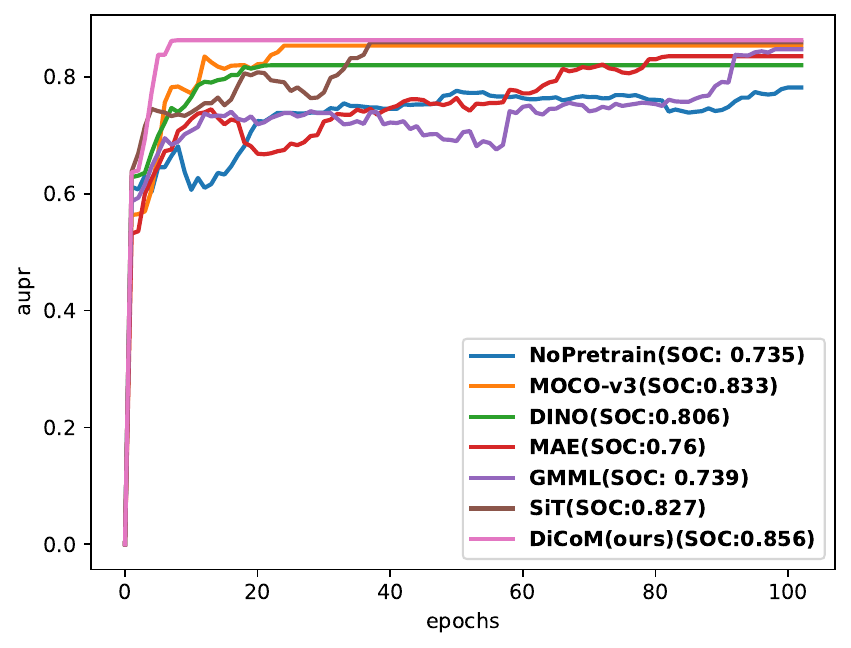}
    \abovecaptionskip=1pt
    \caption{\textbf{Speed of Convergence (SoC)} for various methods. Higher SoC indicates better performance in low resource settings.\vspace{-0.2cm}}
    \label{fig:soc}
\end{figure}

To quantify this, we gauge the speed of convergence (SoC) by defining it as the area under the curve of the maximum AUPR versus epoch curve. SoC serves as an indicator of how rapidly the model attains maximum performance. To simulate low data settings, we use the COVID-PEDs dataset. Figure \ref{fig:soc} illustrates that DiCoM exhibits a high SoC score of 0.856. This is 2.3\% higher than MOCO-v3, the nearest competitor, and 11.9\% higher than the model without pre-training. Additionally, it is noteworthy that DiCoM achieves peak performance in 15 epochs, whereas MOCO-v3 requires at least 27 epochs. The accelerated convergence of DiCoM therefore enhances efficiency, making it suitable and preferable not only in resource-constrained settings but also environmentally friendly.

\section{Discussions}
\begin{figure}[ht]
    \centering
    \includegraphics[width=0.85\columnwidth]{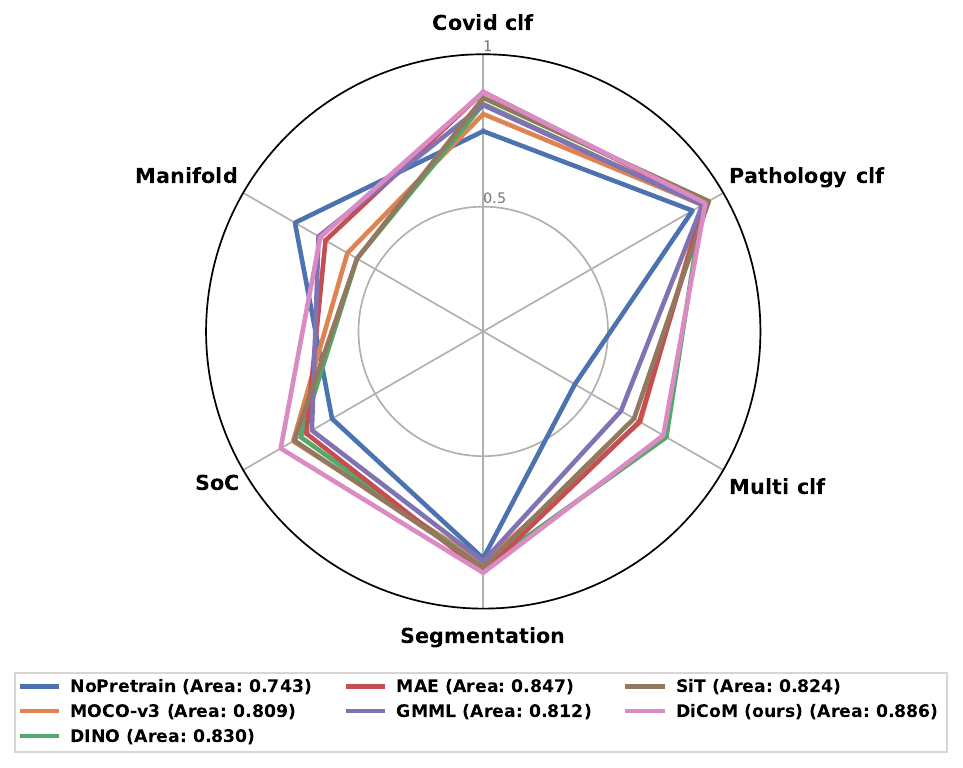}
    \abovecaptionskip=1pt
    \caption{\textbf{Performance summary} of the various pretraining strategies on different tasks. Higher area is an indicator of better overall performance. Abbreviations: clf, classification.\vspace{-0.4cm}}
    \label{fig:overall}
\end{figure}
Our experimental results described in the preceding section support the claim that the representation learning achieved with DiCoM leads to superior performance and generalizability when compared to other pre-training approaches, including fully supervised ViTs, across various domain-specific downstream tasks, as illustrated in Figure \ref{fig:overall}. With area under the curve for DiCoM of 0.886 (highest among compared methods), 
our experimental evaluations affirms that DiCoM excels in generating representations that empowers the ViT model to achieve top performance on not only large-scale datasets but also for datasets with diverse characteristics. These include larger seen distributions during pre-training to smaller unseen out-of-distribution datasets (i.e., pediatric data acquired from different devices and conditions). 
 Furthermore, DiCoM consistently outperforms fully supervised ViT-S models in all classification tasks, achieving a performance gain of $\sim$18.6\%. Our method also exhibits superior performance in segmentation downstream tasks compared to alternative approaches, showing generalizibility in shifts of architecture. In this context, DiCoM achieves a gain of $\sim$4.4\% over fully supervised ViT-S model. 
Moreover, clustering results combining the Rand index and silhouette coefficient validate the superior representation learned by DiCoM. We argue that this clustering superiority could be the reason for its effectiveness in these diverse downstream tasks. Lastly, SoC results reveal that DiCoM's pre-trained weights are helpful for deployment in scenarios with limited data and computational resources, facilitating robust performance with a reduced carbon footprint—a valuable attribute in resource-constrained and environmentally conscious settings.

\section{Conclusions}
\label{sec: conclusions}
Given the importance of CXRs in the current healthcare landscape and the increased care burden within healthcare facilities, we have proposed DiCoM as a paradigm that leverages self-supervised pre-training to address domain-specific downstream tasks related to CXRs.
Looking ahead, we plan to enhance and further extend the generalizibilty of DiCoM paradigm to establish a foundational model for chest radiography by incorporating even more CXR data during pre-training such as MIMIC and other public and private repositories, and including computed tomography scans from chest for volumetric analyses. Such advances would enhance the model's capability for detecting additional lung diseases such as tuberculosis and nodules.

{\small
\bibliographystyle{ieee_fullname}
\bibliography{references}

\begin{thebibliography}{10}\itemsep=-1pt

\bibitem{belarus}
Belarus tuberculosis portal.

\bibitem{abo2021chest}
Sherif~A Abo-Hedibah, Nehal Tharwat, and Ali~H Elmokadem.
\newblock Is chest x-ray severity scoring for covid-19 pneumonia reliable?
\newblock {\em Polish Journal of Radiology}, 86(1):432--439, 2021.

\bibitem{atito2022sb}
Sara Atito, Syed~Muhammad Anwar, Muhammad Awais, and Josef Kittler.
\newblock Sb-ssl: Slice-based self-supervised transformers for knee abnormality classification from mri.
\newblock In {\em Workshop on Medical Image Learning with Limited and Noisy Data}, pages 86--95. Springer, 2022.

\bibitem{atito2021mc}
Sara Atito, Muhammad Awais, Ammarah Farooq, Zhenhua Feng, and Josef Kittler.
\newblock Mc-ssl0. 0: Towards multi-concept self-supervised learning.
\newblock {\em arXiv preprint arXiv:2111.15340}, 2021.

\bibitem{atito2021sit}
Sara Atito, Muhammad Awais, and Josef Kittler.
\newblock Sit: Self-supervised vision transformer.
\newblock {\em arXiv preprint arXiv:2104.03602}, 2021.

\bibitem{atito2022gmml}
Sara Atito, Muhammad Awais, and Josef Kittler.
\newblock Gmml is all you need.
\newblock {\em arXiv preprint arXiv:2205.14986}, 2022.

\bibitem{azad2023foundational}
Bobby Azad, Reza Azad, Sania Eskandari, Afshin Bozorgpour, Amirhossein Kazerouni, Islem Rekik, and Dorit Merhof.
\newblock Foundational models in medical imaging: A comprehensive survey and future vision.
\newblock {\em arXiv preprint arXiv:2310.18689}, 2023.

\bibitem{bachman2019learning}
Philip Bachman, R~Devon Hjelm, and William Buchwalter.
\newblock Learning representations by maximizing mutual information across views.
\newblock {\em Advances in neural information processing systems}, 32, 2019.

\bibitem{balestriero2023cookbook}
Randall Balestriero, Mark Ibrahim, Vlad Sobal, Ari Morcos, Shashank Shekhar, Tom Goldstein, Florian Bordes, Adrien Bardes, Gregoire Mialon, Yuandong Tian, et~al.
\newblock A cookbook of self-supervised learning.
\newblock {\em arXiv preprint arXiv:2304.12210}, 2023.

\bibitem{bao2021beit}
Hangbo Bao, Li Dong, and Furu Wei.
\newblock Beit: Bert pre-training of image transformers.
\newblock {\em arXiv preprint arXiv:2106.08254}, 2021.

\bibitem{caron2021emerging}
Mathilde Caron, Hugo Touvron, Ishan Misra, Herv{\'e} J{\'e}gou, Julien Mairal, Piotr Bojanowski, and Armand Joulin.
\newblock Emerging properties in self-supervised vision transformers.
\newblock In {\em Proceedings of the IEEE/CVF international conference on computer vision}, pages 9650--9660, 2021.

\bibitem{chen2020generative}
Mark Chen, Alec Radford, Rewon Child, Jeffrey Wu, Heewoo Jun, David Luan, and Ilya Sutskever.
\newblock Generative pretraining from pixels.
\newblock In {\em International conference on machine learning}, pages 1691--1703. PMLR, 2020.

\bibitem{chen2020simple}
Ting Chen, Simon Kornblith, Mohammad Norouzi, and Geoffrey Hinton.
\newblock A simple framework for contrastive learning of visual representations.
\newblock In {\em International conference on machine learning}, pages 1597--1607. PMLR, 2020.

\bibitem{das2021comparison}
Karuna~M Das, Jamal~A Alkoteesh, Jumaa Al~Kaabi, Taleb Al~Mansoori, Abbey~J Winant, Rajvir Singh, Rajesh Paraswani, Rizwan Syed, Elsadeg~M Sharif, Ghazala~B Balhaj, et~al.
\newblock Comparison of chest radiography and chest ct for evaluation of pediatric covid-19 pneumonia: Does ct add diagnostic value?
\newblock {\em Pediatric Pulmonology}, 56(6):1409--1418, 2021.

\bibitem{devlin2018bert}
Jacob Devlin, Ming-Wei Chang, Kenton Lee, and Kristina Toutanova.
\newblock Bert: Pre-training of deep bidirectional transformers for language understanding.
\newblock {\em arXiv preprint arXiv:1810.04805}, 2018.

\bibitem{dosovitskiy2020image}
Alexey Dosovitskiy, Lucas Beyer, Alexander Kolesnikov, Dirk Weissenborn, Xiaohua Zhai, Thomas Unterthiner, Mostafa Dehghani, Matthias Minderer, Georg Heigold, Sylvain Gelly, et~al.
\newblock An image is worth 16x16 words: Transformers for image recognition at scale.
\newblock {\em arXiv preprint arXiv:2010.11929}, 2020.

\bibitem{unifesp}
Felipe~Kitamura Eduardo~Farina.
\newblock Unifesp x-ray body part classifier competition, 2022.

\bibitem{gidaris2018unsupervised}
Spyros Gidaris, Praveer Singh, and Nikos Komodakis.
\newblock Unsupervised representation learning by predicting image rotations.
\newblock {\em arXiv preprint arXiv:1803.07728}, 2018.

\bibitem{grill2020bootstrap}
Jean-Bastien Grill, Florian Strub, Florent Altch{\'e}, Corentin Tallec, Pierre~H Richemond, Elena Buchatskaya, Carl Doersch, Bernardo~Avila Pires, Zhaohan~Daniel Guo, Mohammad~Gheshlaghi Azar, et~al.
\newblock Bootstrap your own latent: A new approach to self-supervised learning.
\newblock {\em arXiv preprint arXiv:2006.07733}, 2020.

\bibitem{gu2021domain}
Yu Gu, Robert Tinn, Hao Cheng, Michael Lucas, Naoto Usuyama, Xiaodong Liu, Tristan Naumann, Jianfeng Gao, and Hoifung Poon.
\newblock Domain-specific language model pretraining for biomedical natural language processing.
\newblock {\em ACM Transactions on Computing for Healthcare (HEALTH)}, 3(1):1--23, 2021.

\bibitem{han2022survey}
Kai Han, Yunhe Wang, Hanting Chen, Xinghao Chen, Jianyuan Guo, Zhenhua Liu, Yehui Tang, An Xiao, Chunjing Xu, Yixing Xu, et~al.
\newblock A survey on vision transformer.
\newblock {\em IEEE transactions on pattern analysis and machine intelligence}, 2022.

\bibitem{unetr}
Ali Hatamizadeh, Yucheng Tang, Vishwesh Nath, Dong Yang, Andriy Myronenko, Bennett Landman, Holger~R Roth, and Daguang Xu.
\newblock Unetr: Transformers for 3d medical image segmentation.
\newblock In {\em Proceedings of the IEEE/CVF Winter Conference on Applications of Computer Vision}, pages 574--584, 2022.

\bibitem{Kaiming2021mae}
Kaiming He, Xinlei Chen, Saining Xie, Yanghao Li, Piotr Dollár, and Ross Girshick.
\newblock Masked autoencoders are scalable vision learners.
\newblock {\em arXiv preprint arXiv:2111.06377}, 2021.

\bibitem{he2020momentum}
Kaiming He, Haoqi Fan, Yuxin Wu, Saining Xie, and Ross Girshick.
\newblock Momentum contrast for unsupervised visual representation learning.
\newblock In {\em Proceedings of the IEEE/CVF Conference on Computer Vision and Pattern Recognition}, pages 9729--9738, 2020.

\bibitem{hinton2015distilling}
Geoffrey Hinton, Oriol Vinyals, and Jeff Dean.
\newblock Distilling the knowledge in a neural network.
\newblock {\em arXiv preprint arXiv:1503.02531}, 2015.

\bibitem{hjelm2018learning}
R~Devon Hjelm, Alex Fedorov, Samuel Lavoie-Marchildon, Karan Grewal, Phil Bachman, Adam Trischler, and Yoshua Bengio.
\newblock Learning deep representations by mutual information estimation and maximization.
\newblock In {\em International Conference on Learning Representations (ICLR)}, 2019.

\bibitem{Irvin2019}
Jeremy Irvin, Pranav Rajpurkar, Michael Ko, Yifan Yu, Silviana Ciurea-Ilcus, Chris Chute, Henrik Marklund, Behzad Haghgoo, Robyn Ball, Katie Shpanskaya, Jayne Seekins, David~A. Mong, Safwan~S. Halabi, Jesse~K. Sandberg, Ricky Jones, David~B. Larson, Curtis~P. Langlotz, Bhavik~N. Patel, Matthew~P. Lungren, and Andrew~Y. Ng.
\newblock Chexpert: A large chest radiograph dataset with uncertainty labels and expert comparison.
\newblock {\em 33rd AAAI Conference on Artificial Intelligence, AAAI 2019, 31st Innovative Applications of Artificial Intelligence Conference, IAAI 2019 and the 9th AAAI Symposium on Educational Advances in Artificial Intelligence, EAAI 2019}, pages 590--597, 1 2019.

\bibitem{Jaeger2014}
Stefan Jaeger, Sema Candemir, Sameer Antani, Yì-Xiáng~J. Wáng, Pu-Xuan Lu, and George Thoma.
\newblock Two public chest x-ray datasets for computer-aided screening of pulmonary diseases.
\newblock {\em Quantitative Imaging in Medicine and Surgery}, 4:475, 12 2014.

\bibitem{Johnson2019physionet}
A. Johnson, M. Lungren, Y. Peng, Z. Lu, R. Mark, S. Berkowitz, and S. Horng.
\newblock {MIMIC-CXR-JPG - chest radiographs with structured labels (version 2.0.0)}, 2019.

\bibitem{Johnson2019arxiv}
Alistair E.~W. Johnson, Tom~J. Pollard, Nathaniel~R. Greenbaum, Matthew~P. Lungren, Chih ying Deng, Yifan Peng, Zhiyong Lu, Roger~G. Mark, Seth~J. Berkowitz, and Steven Horng.
\newblock {MIMIC-CXR-JPG, a large publicly available database of labeled chest radiographs}.
\newblock {\em arXiv preprint arXiv:1901.07042}, 1 2019.

\bibitem{kendall2018multi}
Alex Kendall, Yarin Gal, and Roberto Cipolla.
\newblock Multi-task learning using uncertainty to weigh losses for scene geometry and semantics.
\newblock In {\em Proceedings of the IEEE conference on computer vision and pattern recognition}, pages 7482--7491, 2018.

\bibitem{Kermany2018}
Daniel~S. Kermany, Michael Goldbaum, Wenjia Cai, Carolina~C.S. Valentim, Huiying Liang, Sally~L. Baxter, Alex McKeown, Ge Yang, Xiaokang Wu, Fangbing Yan, Justin Dong, Made~K. Prasadha, Jacqueline Pei, Magdalena Ting, Jie Zhu, Christina Li, Sierra Hewett, Jason Dong, Ian Ziyar, Alexander Shi, Runze Zhang, Lianghong Zheng, Rui Hou, William Shi, Xin Fu, Yaou Duan, Viet~A.N. Huu, Cindy Wen, Edward~D. Zhang, Charlotte~L. Zhang, Oulan Li, Xiaobo Wang, Michael~A. Singer, Xiaodong Sun, Jie Xu, Ali Tafreshi, M.~Anthony Lewis, Huimin Xia, and Kang Zhang.
\newblock Identifying medical diagnoses and treatable diseases by image-based deep learning.
\newblock {\em Cell}, 172:1122--1131.e9, 2 2018.

\bibitem{Kitamura2023}
Felipe Kitamura, Lilian Mallagoli, and Paulo Kuriki.
\newblock Spr x-ray gender prediction challenge, 2023.

\bibitem{kramer1991nonlinear}
Mark~A Kramer.
\newblock Nonlinear principal component analysis using autoassociative neural networks.
\newblock {\em AIChE journal}, 37(2):233--243, 1991.

\bibitem{ma2023foundation}
DongAo Ma, Jiaxuan Pang, Michael~B Gotway, and Jianming Liang.
\newblock Foundation ark: Accruing and reusing knowledge for superior and robust performance.
\newblock In {\em International Conference on Medical Image Computing and Computer-Assisted Intervention}, pages 651--662. Springer, 2023.

\bibitem{noroozi2016unsupervised}
Mehdi Noroozi and Paolo Favaro.
\newblock Unsupervised learning of visual representations by solving jigsaw puzzles.
\newblock In {\em European conference on computer vision}, pages 69--84. Springer, 2016.

\bibitem{oord2018representation}
Aaron van~den Oord, Yazhe Li, and Oriol Vinyals.
\newblock Representation learning with contrastive predictive coding.
\newblock {\em arXiv preprint arXiv:1807.03748}, 2018.

\bibitem{parvaiz2022vision}
Arshi Parvaiz, Muhammad~Anwaar Khalid, Rukhsana Zafar, Huma Ameer, Muhammad Ali, and Muhammad~Moazam Fraz.
\newblock Vision transformers in medical computer vision--a contemplative retrospection.
\newblock {\em arXiv preprint arXiv:2203.15269}, 2022.

\bibitem{NEURIPS2020_29539ed9}
Massimiliano Patacchiola and Amos~J Storkey.
\newblock Self-supervised relational reasoning for representation learning.
\newblock In {\em Advances in Neural Information Processing Systems}, volume~33, pages 4003--4014, 2020.

\bibitem{pathak2016context}
Deepak Pathak, Philipp Krahenbuhl, Jeff Donahue, Trevor Darrell, and Alexei~A Efros.
\newblock Context encoders: Feature learning by inpainting.
\newblock In {\em Proceedings of the IEEE conference on computer vision and pattern recognition}, pages 2536--2544, 2016.

\bibitem{pavlova2022covidx}
Maya Pavlova, Tia Tuinstra, Hossein Aboutalebi, Andy Zhao, Hayden Gunraj, and Alexander Wong.
\newblock Covidx cxr-3: A large-scale, open-source benchmark dataset of chest x-ray images for computer-aided covid-19 diagnostics.
\newblock {\em arXiv preprint arXiv:2206.03671}, 2022.

\bibitem{unet}
Olaf Ronneberger, Philipp Fischer, and Thomas Brox.
\newblock U-net: Convolutional networks for biomedical image segmentation.
\newblock In {\em International Conference on Medical image computing and computer-assisted intervention}, pages 234--241. Springer, 2015.

\bibitem{sriram2021covid}
Anuroop Sriram, Matthew Muckley, Koustuv Sinha, Farah Shamout, Joelle Pineau, Krzysztof~J Geras, Lea Azour, Yindalon Aphinyanaphongs, Nafissa Yakubova, and William Moore.
\newblock Covid-19 prognosis via self-supervised representation learning and multi-image prediction.
\newblock {\em arXiv preprint arXiv:2101.04909}, 2021.

\bibitem{pmlr-v139-touvron21a}
Hugo Touvron, Matthieu Cord, Matthijs Douze, Francisco Massa, Alexandre Sablayrolles, and Herve Jegou.
\newblock Training data-efficient image transformers \& distillation through attention.
\newblock In {\em International Conference on Machine Learning}, volume 139, pages 10347--10357, July 2021.

\bibitem{trinh2019selfie}
Trieu~H Trinh, Minh-Thang Luong, and Quoc~V Le.
\newblock Selfie: Self-supervised pretraining for image embedding.
\newblock {\em arXiv preprint arXiv:1906.02940}, 2019.

\bibitem{UGALDE20211039}
Irma~T. Ugalde, Samuel Prater, Marylou Cardenas-Turanzas, Nipa Sanghani, Donna Mendez, John Peacock, Grace Guvernator, Christine Koerner, and Myron Allukian.
\newblock Chest x-ray vs. computed tomography of the chest in pediatric blunt trauma.
\newblock {\em Journal of Pediatric Surgery}, 56(5):1039--1046, 2021.

\bibitem{vincent2008extracting}
Pascal Vincent, Hugo Larochelle, Yoshua Bengio, and Pierre-Antoine Manzagol.
\newblock Extracting and composing robust features with denoising autoencoders.
\newblock In {\em Proceedings of the 25th international conference on Machine learning}, pages 1096--1103, 2008.

\bibitem{Covidnet}
Linda Wang, Zhong~Qiu Lin, and Alexander Wong.
\newblock Covid-net: a tailored deep convolutional neural network design for detection of covid-19 cases from chest x-ray images.
\newblock {\em Scientific Reports}, 10(1):19549, Nov 2020.

\bibitem{Wang2017}
Xiaosong Wang, Yifan Peng, Le Lu, Zhiyong Lu, Mohammadhadi Bagheri, and Ronald~M Summers.
\newblock Chestx-ray8: Hospital-scale chest x-ray database and benchmarks on weakly-supervised classification and localization of common thorax diseases.
\newblock pages 2097--2106, 2017.

\bibitem{wu2018unsupervised}
Zhirong Wu, Yuanjun Xiong, Stella~X Yu, and Dahua Lin.
\newblock Unsupervised feature learning via non-parametric instance discrimination.
\newblock In {\em Proceedings of the IEEE conference on computer vision and pattern recognition}, pages 3733--3742, 2018.

\bibitem{xie2022simmim}
Zhenda Xie, Zheng Zhang, Yue Cao, Yutong Lin, Jianmin Bao, Zhuliang Yao, Qi Dai, and Han Hu.
\newblock Simmim: A simple framework for masked image modeling.
\newblock In {\em Proceedings of the IEEE/CVF Conference on Computer Vision and Pattern Recognition}, pages 9653--9663, 2022.

\bibitem{zhang2016colorful}
Richard Zhang, Phillip Isola, and Alexei~A Efros.
\newblock Colorful image colorization.
\newblock In {\em European conference on computer vision}, pages 649--666. Springer, 2016.

\end{thebibliography}
}
\end{document}